\documentclass[a4paper]{article}

\usepackage{atmohead2013}
\usepackage[english]{babel}

\title{Atmospheric Aerosol Attenuation Measurements at
the Pierre Auger \\
Observatory}

\shorttitle{Aerosol Measurements at the Pierre Auger Observatory}

\authors{
Laura Valore$^{1}$,
for the Pierre Auger Collaboration$^{2}$.
}

\afiliations{
$^1$ University of Naples and INFN Naples \\
$^2$ Full author list: http://www.auger.org/archive/authors\_2013\_05.html\\
}

\email{valore@na.infn.it}

\abstract{
The Fluorescence Detector (FD) of the Pierre Auger Observatory provides a nearly calorimetric measurement 
of the primary particle energy, since the fluorescence light produced is proportional to the energy dissipated 
by an Extensive Air Shower (EAS) in the atmosphere. The atmosphere therefore acts as a giant calorimeter, 
whose properties need to be well known during data taking. Aerosols play a key role in this scenario, since their effect on light transmission is highly variable even on a time scale of one hour, and the corresponding 
correction to EAS energy can range from a few percent to more than 40$\%$. For this reason, hourly Vertical 
Aerosol Optical Depth ($\rm{\tau_{aer}(h)}$) profiles are provided for each of the four FD stations. 
Starting from 2004, up to now 9 years of $\rm{\tau_{aer}(h)}$ profiles have been produced using data from 
the Central Laser Facility (CLF) and the eXtreme Laser Facility (XLF) of the Pierre Auger Observatory. 
The two laser facilities, the techniques developed to measure 
$\rm{\tau_{aer}(h)}$ profiles using laser data and the results will be discussed.}


\begin{document}
\maketitle
\section{Introduction}
Ultra High Energy Cosmic Rays (UHECR, $\rm E>10^{18} eV$) entering the atmosphere 
cannot be directly detected due to their extremely low flux.  For this reason, the 
properties of primary particles (energy, mass composition, direction) are deduced
from the study of the cascade of secondary particles (Extensive Air Showers, EAS) 
that originates in the atmosphere due to the interaction of those primaries with 
air molecules. The Pierre Auger Observatory is the largest detector of EAS ever built, 
covering an area of $\rm 3000$ $\rm km^2$, located in Argentina in the province of Mendoza. The 
observatory uses two techniques at the same time : the detection of particles at ground
level with the Surface Detector (SD) and the observation of the longitudinal 
development of the EAS by detecting the fluorescence light emitted with the Fluorescence 
Detector (FD). The SD array is composed of more than 1600 water Cherenkov detectors,
overlooked by 27 fluorescence telescopes grouped in 4 sites located at 
the array periphery. The observatory was completed in 2008. 
The FD is designed to perform a nearly calorimetric measurement of 
the energy of cosmic ray primaries: the detected flux of fluorescence photons, emitted by nitrogen 
air molecules excited by EAS charged particles, is proportional to the energy deposit per unit 
slant depth of the traversed atmosphere. Due to the constantly changing properties of the calorimeter 
(i.e. the atmosphere), in which the light is both produced and through which it is 
transmitted, an extensive system with several instruments has been set up to perform a 
continuous monitoring of its properties. In particular, the aerosol attenuation of the 
fluorescence light, highly variable on a time scale of one hour, needs to be constantly
measured during data acquisition. If the aerosol attenuation is not taken into account, the shower 
energy reconstruction is biased by 8 to 25\% in the energy range measured by the Pierre Auger 
Observatory. On average, 20\% of all showers have an energy correction larger than 20\%, 7\% of 
showers are corrected by more than 30\% and 3\% of showers are corrected by more than 40\% \cite{bib:SegevPaper}. 
At the Pierre Auger Observatory, hourly vertical aerosol optical depth profiles 
have been produced for each FD site from January 2004 to December 2012 for a correct reconstruction of FD events.

\section{The Aerosol Monitoring System}
\begin{figure}[h]
  \centering
  \includegraphics[width=0.3\textwidth]{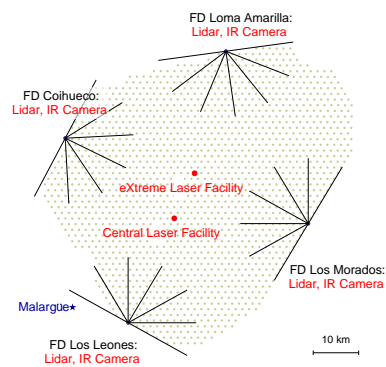}
  \caption{Map of the Pierre Auger Observatory aerosol monitoring system.}
  \label{fig:map}
 \end{figure}

The Pierre Auger Observatory has a diverse atmospheric monitoring system; among its instruments, many are
dedicated to aerosol attenuation measurements. A map of the observatory, with the main aerosol
monitoring devices, is shown in figure \ref{fig:map}. 

Two laser facilities, the Central Laser Facility (CLF) \cite{bib:CLF} and the eXtreme Laser Facility (XLF), produce laser
beams each from a position nearly equidistant from three out of four FD sites. The CLF was built in late 2003 
and has been operational from early 2004, while the XLF was built later and has been operational since 2011. 
Among the many uses of these test beams, the analysis of laser data permits one to evaluate the aerosol attenuation: 
once the nominal energy of the laser source is known, the number of photons reaching the FD depends on the 
properties of the atmosphere, therefore aerosol attenuation can be inferred \cite{bib:CLFpaper}. 
The details of the analyses will be described in the following section. 
Four LIDAR stations, one at each FD site, are equipped with a UV laser source for the detection of the elastic
backscattered light to record local aerosol conditions and clouds \cite{bib:LIDAR}. They can also provide rapid monitoring
 after the detection of high energy showers \cite{bib:rapid}. A Raman LIDAR station was also installed in early 2004 at one of the four 
FD sites. A major upgrade has been completed in the past months at the CLF site, that includes the addition of a Raman LIDAR to the system to 
perform $\rm \tau_{aer}(h)$ independent measurements. 
 
\subsection{The Laser Facilities}
\begin{figure}[h]
  \centering
  \includegraphics[width=0.3\textwidth]{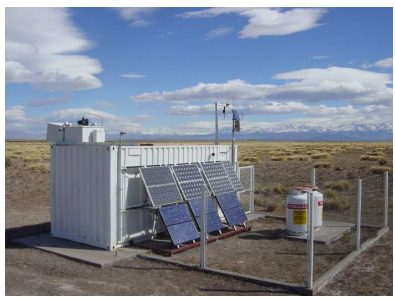}
  \caption{A picture of the CLF.}
  \label{fig:clfphoto}
 \end{figure}

The CLF and the XLF use a frequency tripled Nd:YAG laser, control hardware and optics to direct a calibrated pulsed
UV beam into the sky. Its wavelength of 355 nm is near the center of the nitrogen fluorescence spectrum.
In figure \ref{fig:clfphoto}, a picture of the CLF is shown. The two laser facilities are solar-powered and operated remotely
during FD shifts.

The laser is mounted on an optical table that also houses most of the other
optical components. Two selectable beam configurations --~vertical and steerable~-- are available.
The inclined laser shots can be used to calibrate the pointing and time offsets of the fluorescence
telescopes; for the aerosol attenuation techniques described here, only the vertical beam is used.
The Nd:YAG laser emits linearly polarized light: a depolarizer is used to randomly polarize the light
so that equal amounts of light are scattered in the azimuthal directions of each FD site.
The laser energy of the CLF is monitored by a pyroelectric probe receiving a fraction of the laser
beam for a relative calibration of each laser shot. Additionally, absolute calibrations are performed
periodically, capturing the entire laser beam with an external radiometer before sending the 
laser light to the sky. The periodic absolute calibration permits one to correct the sky energy
for the effects related to dust accumulation on some of the optics of the laser bench. 
The XLF is equipped with a combined system of a pick-off probe for relative calibration
and an automated calibration system which performs absolute calibrations on a nightly basis using
a robotic arm moving a calibration probe in the beam path of the XLF laser.

\begin{figure}[h]
  \centering
  \includegraphics[width=0.25\textwidth]{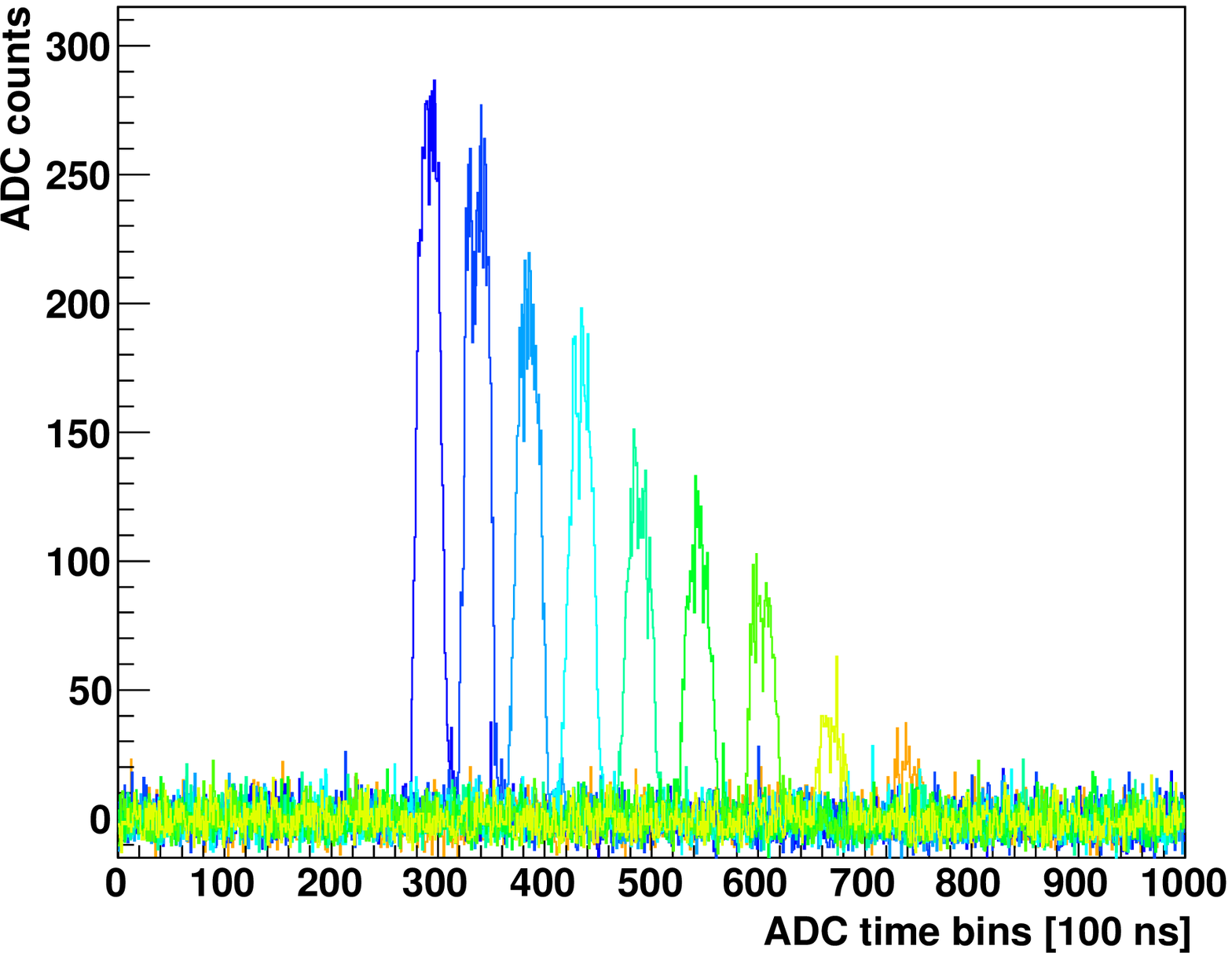}%
  \includegraphics[width=0.25\textwidth]{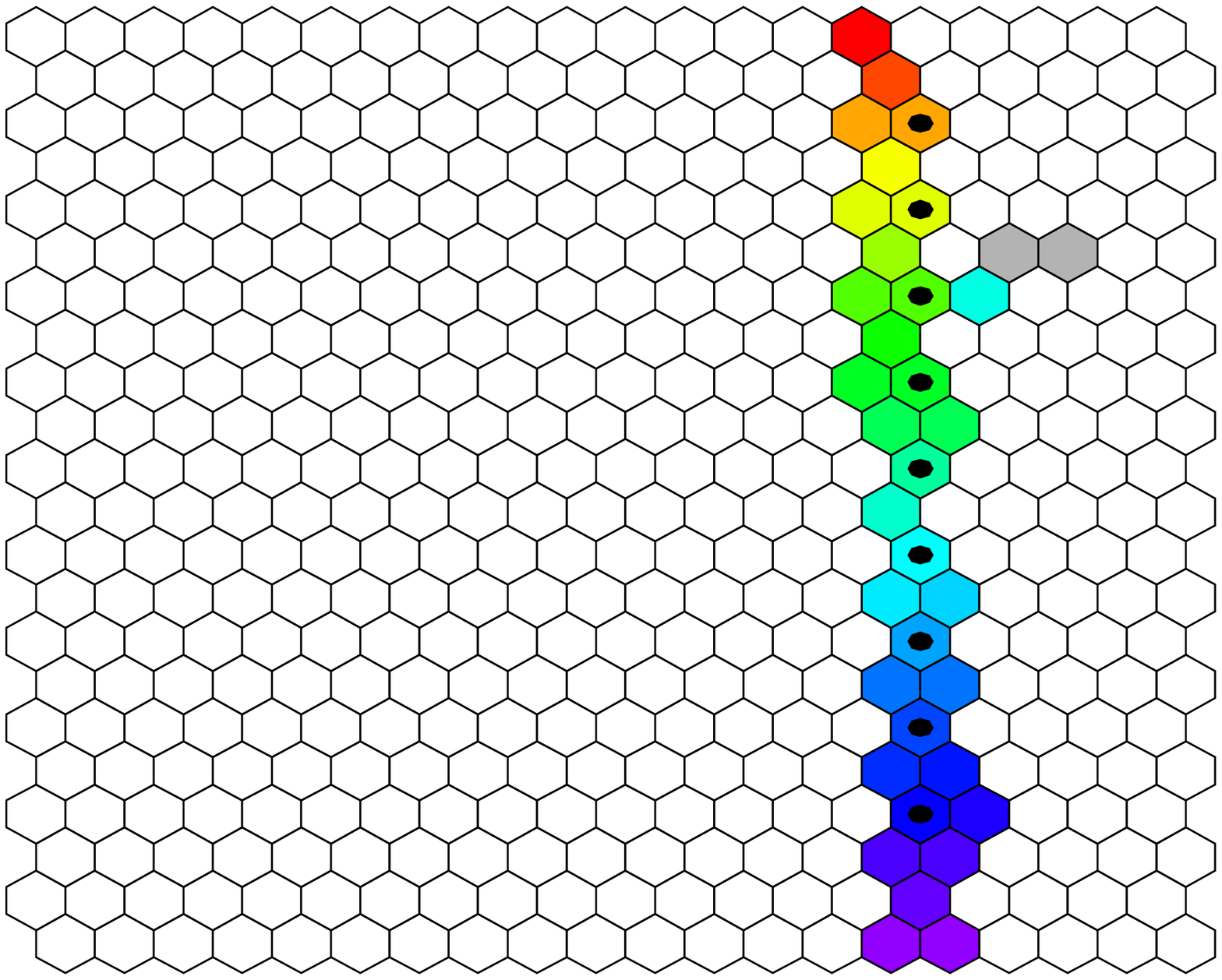}
  \caption{A 7~mJ CLF vertical event as recorded by the FD. 
    Left panel: ADC counts vs.\ time (100~ns bins). Displayed data
      are for the marked pixels in the right panel. Right panel: Camera trace.
      The color code indicates the sequence in which the pixels were triggered (from blue to red).}
  \label{fig:clf_fdeyedisplay}
 \end{figure}

The CLF and the XLF fire 50~vertical shots at 0.5~Hz repetition rate every 15~minutes
during the FD data acquisition. The light scattered out of the laser beam is recorded by the FD. 
Laser tracks are recorded by the telescopes in the same format used for air shower
measurements.  Specific GPS timing is used to distinguish laser
from air shower events. The direction, time, and relative energy of each laser
pulse is recorded at the laser facility and later matched to the corresponding laser event
in the FD data. In figure \ref{fig:clf_fdeyedisplay}, a single 7 mJ CLF vertical
shot as recorded from the Los Leones FD site is shown. 

\section{Analysis Techniques}

Two independent analyses have been developed to measure hourly aerosol attenuation in
the FD field of view using vertical CLF and XLF shots : the Data Normalized Analysis and 
the Laser Simulation Analysis \cite{bib:CLFpaper}. Both methods are based on the idea that 
laser light is attenuated in the same way as fluorescence light as it propagates towards the
FD. Therefore, the analysis of the amount of laser light that reaches the FD as a function 
of time can be used to infer the attenuation due to aerosols between the position of the laser 
and each FD building. In detail :
 
\begin{itemize}
\item The Data Normalized Analysis (DN) is based on the comparison of measured laser profiles with a reference 
clear night profile in which the light attenuation is dominated by molecular scattering.
\item The Laser Simulation Analysis (LS) is based on the comparison of measured laser light profiles to
 simulations generated in various atmospheres in which the aerosol attenuation is described
 by a parametric model.
\end{itemize}

Using measurements recorded on extremely clear nights where molecular Rayleigh scattering dominates,
laser observations can be normalized without the need for absolute photometric calibrations 
of the FD or laser. These ``reference clear nights'' are identified using a procedure looking for
profiles with maximum photon transmission and maximum compatibility with the shape of a profile
simulated in conditions with negligible aerosol attenuation. The level of compatibility between simulated and
measured clear night profiles is established by means of a Kolmogorov-Smirnov test. The procedure is repeated
for each FD site, for each laser facility. One reference clear night per year is selected. Cross checks with 
another instrument, the Aerosol Phase Function Monitor \cite{bib:APF}, are performed to confirm the 
validity of the chosen reference profiles. 
To minimize fluctuations, both analyses make use of average light profiles measured at the 
aperture of the FD buildings normalized to a fixed reference energy.  

\subsection{Data Normalized Analysis}
\begin{figure}[h]
  \centering
  \includegraphics[width=0.5\textwidth]{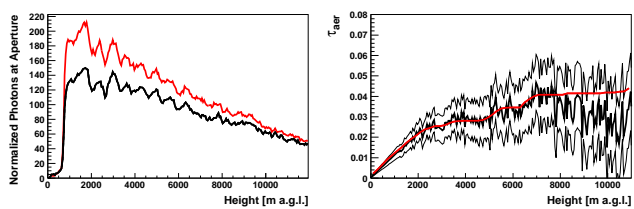}
  \caption{In black the measured light profile, in red the reference one,
and corresponding  $\rm\tau_{aer}(h)$ as measured using the Data Normalized Analysis 
in average conditions.}
  \label{fig:DN}
 \end{figure}

The Data Normalized Analysis is an iterative procedure that compares hourly average profiles 
to reference clear night profiles. The first step is to build the 4 quarter-hour 50 shots profiles, 
normalized to 1 mJ. During this procedure, clouds positioned above the vertical laser beam
are marked by comparing the photon transmission of the quarter-hour profile to that of the clear profile. 
The ratio $\rm T_{quarter}/T_{clean}$ is used to identify clouds and set the minimum cloud layer altitude. 
Hours are marked as cloudy only if clouds are found in at least two quarter-hour sets. 
After cloud identification, the full hour profile is built averaging all quarter-hour profiles available. 
Assuming that the atmosphere is horizontally uniform, the Vertical Aerosol Optical Depth $\rm\tau_{aer}^{DN}(h)$ 
is measured as
$$
 \rm\tau_{aer}^{DN}(h) = \frac{ln \rm N_{mol}(h) - ln N_{obs}(h)}{1 + cosec(\theta)}
$$
where $\rm N_{mol}$(h) is the number of photons from the reference clear profile as a function of height,
$\rm N_{obs}(h)$ is the number of photons from the observed hourly profile as a function of height and 
$\theta$ is the elevation angle on the camera of each laser track segment. 
This calculation does not take into account the scattering of the laser beam itself due to aerosols. 
To overcome this, $\rm\tau_{aer}^{DN}(h)$ is differentiated to calculate the aerosol
extinction coefficient $\rm \alpha(h)$ over short intervals in which the aerosol scattering conditions change 
slowly. The final $\rm\tau_{aer}^{DN}(h)$ is estimated by re-integrating $\rm \alpha(h)$ (figure \ref{fig:DN}).
The aerosol attenuation profile is calculated from the FD site altitude up to the cloud lower layer height or the
highest point in the FD field of view.

\subsection{Laser Simulation Analysis}
\begin{figure}[h]
  \centering
  \includegraphics[width=0.25\textwidth]{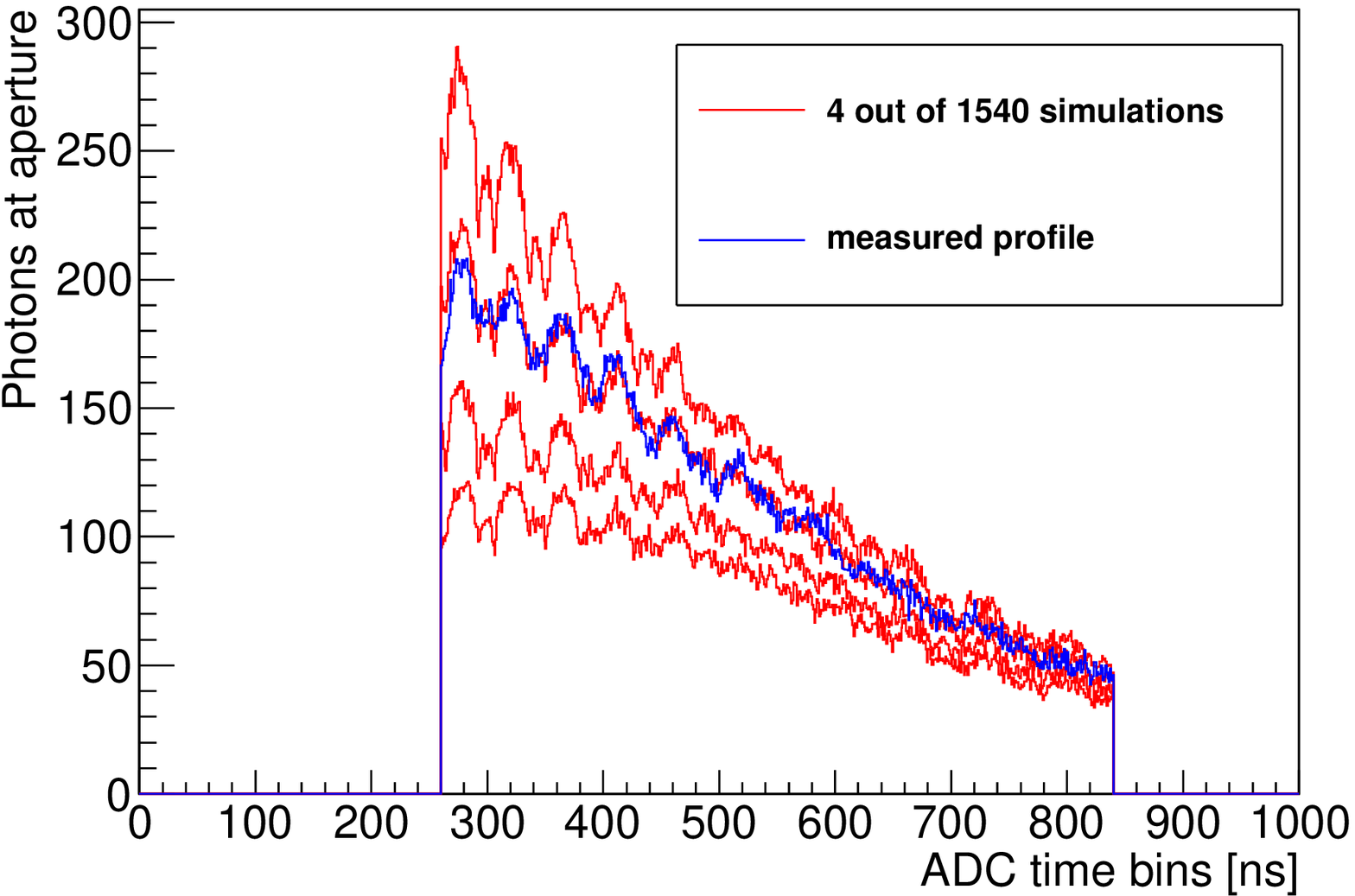}%
  \includegraphics[width=0.25\textwidth]{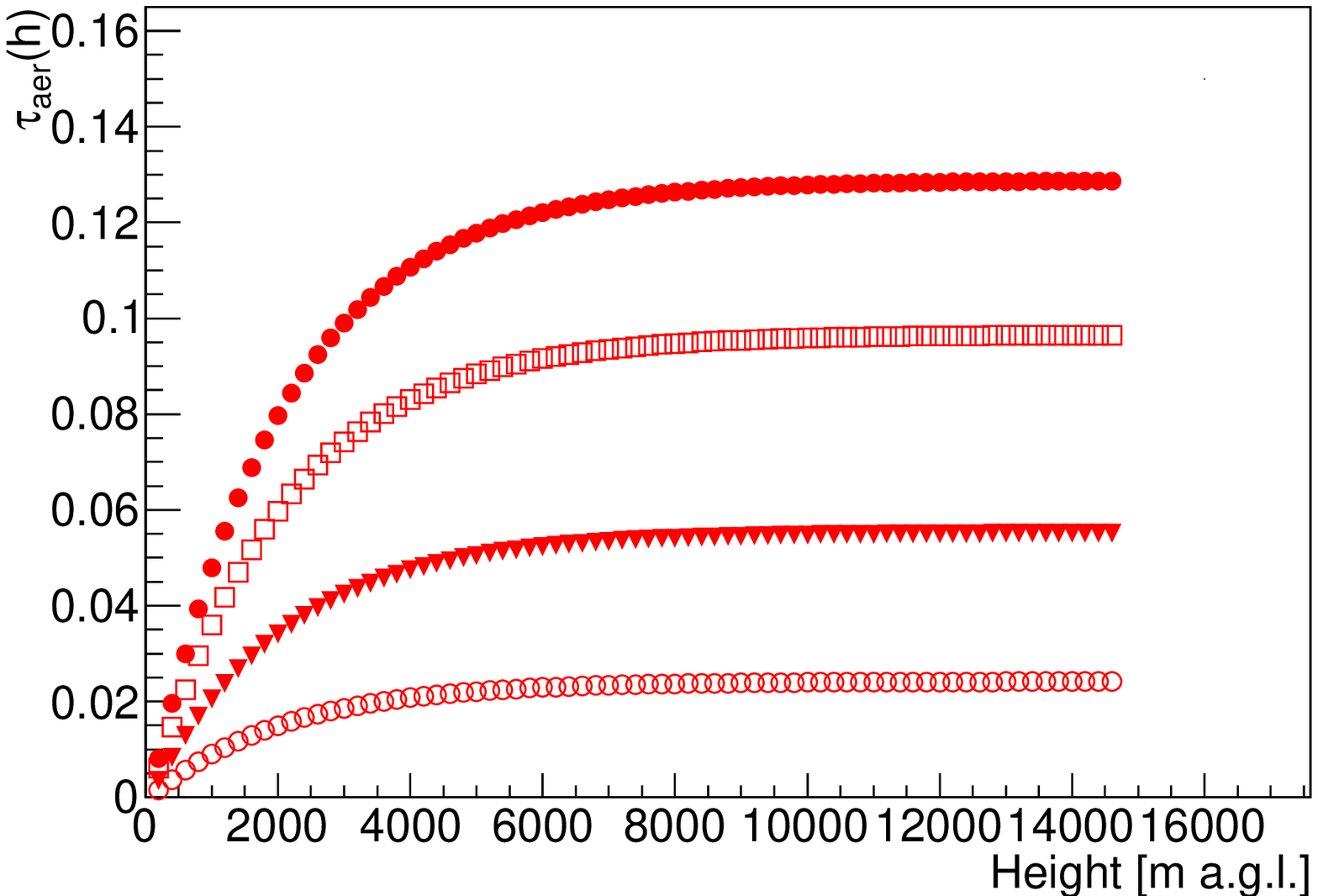}
  \caption{Left : four out of the 1540 simulated profiles of a monthly grid (red), superimposed on a measured profile (blue). 
    Right : the four $\rm\tau_{aer}^{LS}(h)$ profiles corresponding to the simulated CLF profiles.}
  \label{fig:LS}
 \end{figure}

The Laser Simulation Analysis is based on the comparison of average 50-shots-quarter-hour light profiles 
normalized at 6.5 mJ to simulations generated at the same energy, fixing the initial number of
photons emitted by the simulated vertical laser source.  While energy and geometry of the simulated laser 
event are fixed, the atmospheric conditions, defined by aerosol and air density profiles, are variable 
and described by means of a two-parameter models: the aerosol horizontal attenuation length $\rm L_{aer}$ 
and the aerosol scale height $\rm H_{aer}$.
The former describes the horizontal light attenuation due to aerosols at ground level, the latter accounts for
its dependence on the height. With this parameterization, the expression of the vertical aerosol optical 
depth $\rm\tau_{aer}^{LS}(h)$ between points at altitude $\rm h_1$ and $\rm h_2$ is :
$$
\rm \tau_{aer}^{LS}(h_2 - h_1) = -\frac{H_{aer}}{L_{aer}} \left[ \exp{ \left( -\frac{h_2}{H_{\rm aer} }\right) }
    - \exp{ \left( - \frac{h_1}{H_{\rm aer}} \right) }  \right] 
$$
For this analysis, the grid is generated by varying $\rm L_{aer}$ from 5 to 50~km in steps 
of 1.25 km and from 50 to 150 km in steps of 2.5~km, and varying $\rm H_{aer}$ from 0.25 km to 5~km
in steps of 0.25~km. This corresponds to a total of 1540~profiles. An average monthly description 
of air density profiles, measured at the observatory site, as a result of an intense campaign of radiosonde
measurements, is used for the simulation. A total of 13452 profiles are simulated to
reproduce the wide range of possible atmospheric conditions on site.
Each measured profile is compared to the grid and the 
simulated profile closest to the measured event is identified and its associated parameters are used to
calculate $\rm\tau_{aer}^{LS}(h)$ (figure \ref{fig:LS}). 
The quantification of the difference between measured and simulated
profiles and the method to identify the closest simulation are the crucial 
points of this analysis. After validation tests on simulations of different 
methods, the pair $\rm L_{aer}^{\rm{best}}$ and 
$\rm H_{aer}^{\rm{best}}$ chosen is the one that
minimizes the square difference $D^2$ between measured and simulated profiles
computed for each bin, where $D^2 = [\sum_i(\Phi_i^{\rm meas} - \Phi_i^{\rm
sim})^2]$ and $\Phi_i$ are reconstructed photon numbers at the FD aperture in
each time bin.

During the procedure, clouds are identified and the aerosol attenuation profile
is measured up to the cloud lower layer height. 
Clouds are identified working on the profile of the difference between the measured and 
best simulated profile. With this choice, the baseline is close to zero and the peaks and 
holes in the signal are clearly visible. The signal to noise ratio and the highest/lowest 
signal are used to mark the cloud lowest layer height, and to assign the maximum height of the 
aerosol profile measured.

The Laser Simulation Analysis extrapolates the aerosol attenuation for each
quarter hour CLF profile; then the four measured aerosol profiles are averaged
to obtain the hourly information needed for the air shower reconstruction. 

\subsection{Uncertainties}
Various uncertainties were identified in the methods for the determination of $\rm \tau_{aer}(h)$ profiles.
The uncertainties are separated into systematic and statistical contributions. 
These assignments were based on whether the effect of the uncertainty would be 
correlated over the EAS data sample, or would be largely uncorrelated from one EAS to the next (see table \ref{tab:errors}). 

\begin{table}[h]
\begin{center}
\begin{tabular}{|l|c|c|}
\hline  & Correlated & Uncorrelated \\ \hline
Relative FD Calibration        & 2\%        & 4\% \\ \hline
Relative Laser Energy (CLF)    & 1--2.5\%  & 2\% \\ \hline
Relative Laser Energy (XLF)    & 1\%        & 2\% \\ \hline
Reference Clear Night          & 3\%        & -   \\ \hline
Atmospheric Fluctuations       & -          & $\sim3\%$ \\ \hline
\end{tabular}
\caption{Uncertainties in the determination of $\rm \tau_{aer}(h)$.}
\label{tab:errors}
\end{center}
\end{table}

The two analysis techniques described make use of ratios of FD events and are therefore not sensitive 
to the absolute photometric calibration of either the laser or the FD. As a consequence, the calibration 
correlated uncertainties in table \ref{tab:errors} are those that describe how accurately drifts in the FD 
and laser energy calibrations were tracked over the period between reference nights.  
For the CLF, the 1-2.5\% value corresponds to different epochs over the 10 year life of the system and 
depends on how well the effect of dust accumulation on the optics downstream of the monitor probe was tracked.  
The corresponding term for the XLF (1\%) is lower due to the automated calibration system that tracks beam 
energy and polarization.
The uncorrelated error of the relative FD calibration was estimated to be 4\%, and includes an estimate of the 
variability in FD calibration during the night.  
A 3\% correlated uncertainty was estimated as due to the choice of the reference clear night. 
Finally the uncorrelated error due to the atmospheric fluctuations within the hour is about 3\%. 
These uncertainties are estimated separately for each of the two analyses described. 
In the Laser Simulation Analysis a 2\% uncorrelated 
uncertainty is added to take into account how well the parametric model used describes the real aerosol attenuation conditions.
\section{9 years of aerosol attenuation profiles}
\begin{figure}[h]
  \centering
  \includegraphics[width=0.3\textwidth]{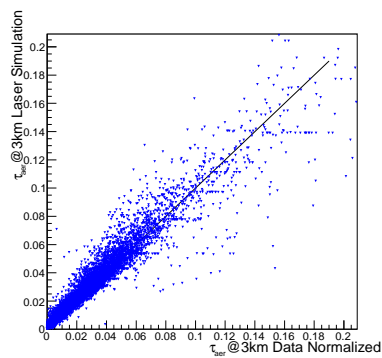}
  \caption{Comparison of $\rm \tau_{aer}(h)$ at 3 km above ground measured with the two analyses. 
9 years of data are shown. }
  \label{fig:corr}
 \end{figure}

The hourly aerosol attenuation profiles over 9 years (from January 2004 to December 2012) 
have been measured using the two analyses described. The compatibility of the results is
evident, as in figure \ref{fig:corr}, where the correlation of $\rm \tau_{aer}^{DN}$  
versus $\rm \tau_{aer}^{LS}$ measured at 3 km is shown. 
The spread of the points is within error bands. 

\begin{figure}[h]
  \centering
  \includegraphics[width=0.4\textwidth]{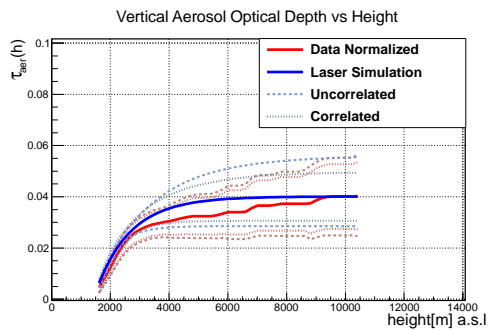}
  \caption{Hourly aerosol profiles measured with the Data Normalized (red) and Laser Simulation (blue) 
analyses in average conditions. Uncertainties are shown.}
  \label{fig:comp_prof}
 \end{figure}

Hourly profiles measured with the two analyses together with correlated and uncorrelated error
bands in average aerosol attenuation conditions are shown in figure \ref{fig:comp_prof}.

\begin{figure}[h]
  \centering
  \includegraphics[width=0.5\textwidth]{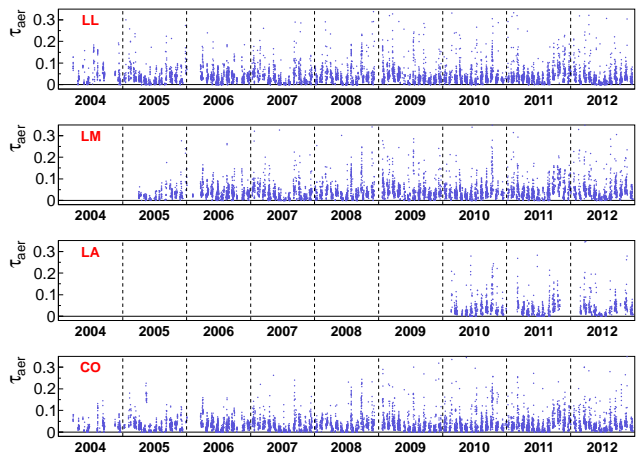}
  \caption{9 years of $\rm \tau_{aer}$ measured at 3 km above ground.}
  \label{fig:vaod_vs_time}
 \end{figure}

The measured aerosol profiles are stored in the aerosol attenuation database of the 
Pierre Auger Observatory for the reconstruction of EAS data. Due to its location, XLF events 
are used to produce aerosol profiles for Loma Amarilla and CLF events are used for Los Leones, 
Los Morados and Coihueco. The database is filled with results obtained with the Data Normalized 
analysis, while results from Laser Simulation analysis are used to fill gaps. A total of 10430 hours
are stored in the aerosol database for the Los Leones site, 9302 for Los Morados, 2270 for Loma Amarilla and
10430 for Coihueco. In figure \ref{fig:vaod_vs_time}, $\rm \tau_{aer}$ measured at 3 km above ground as a 
function of time is shown for each FD site. 
The seasonal variation of  $\rm \tau_{aer}$ is visible: every year, lower values 
are measured in austral winter with respect to summer. 

\end{document}